# Magnesium-graphene interphase boundaries created by high-pressure torsion enhance hydrogen storage kinetics: Mechanisms and significance of activation energy and frequency factor

Runchen Zhou[1,2], Payam Edalati[3], Anthony Alhayek[4,5], Shivam Dangwal[1,2], Marc Novelli[4,5], Md. Amirul Islam[1], Baran Bidyut Saha[1], Thierry Grosdidier[4,5] and Kaveh Edalati[1,2,*]

[1] WPI, International Institute for Carbon-Neutral Energy Research (WPI-I2CNER), Kyushu University, Fukuoka 819-0395, Japan
[2] Department of Automotive Science, Kyushu University, Fukuoka 819-0395, Japan
[3] Faculdade de Engenharia Mecânica (FEM), Universidade Estadual de Campinas (UNICAMP), Limeira, Brazil
[4] Université de Lorraine, Laboratoire d'Etude des Microstructures et de Mécanique des Matériaux (LEM3 UMR 7239), 7 rue Félix Savart, BP 15082, Metz F-57073, France
[5] Université de Lorraine, Laboratory of Excellence on Design of Alloy Metals for low-mass Structures (DAMAS), Metz F-57045, France

A strategy to overcome sluggish hydrogenation/dehydrogenation of magnesium is demonstrated by creating magnesium-graphene interphase boundaries via high-pressure torsion (HPT). HPT reduces the grain size of pure magnesium from ~1 mm to ~850 nm, with 70% of grain boundaries having high misorientation angles. Graphene addition leads to even finer grain sizes of 10–500 nm with a bimodal morphology. The magnesium-graphene composites exhibit superior kinetics at 623 K while maintaining high air resistance. Kinetic modeling reveals that the rate-controlling mechanism transits from interfacial reaction in coarse-grained magnesium to atomic diffusion in magnesium-graphene nanocomposites. Kissinger analysis shows that the activation energy for hydrogen desorption remains unchanged at 145±2 kJ/mol, regardless of the presence of grain or interphase boundaries. However, the frequency factor (number of successful attempts to overcome the activation energy) increases with the generation of interfaces, which serve as sites for hydrogen diffusion and heterogeneous metal/hydride nucleation. These findings highlight the impact of interphase boundary engineering via severe plastic deformation for enhancing the kinetics and air resistance of hydrogen storage materials.

*Corresponding author (email: kaveh.edalati@kyudai.jp)

## 1. Introduction

The international quest for sustainable and green energy solutions has directed researchers to find alternative energy carriers to lower the reliance on fossil fuels. Hydrogen, with a large gravimetric energy density and zero-$CO_2$ emission, has emerged as a favorable candidate [1]. However, developing safe, efficient and compact storage systems is still a vital technological bottleneck for a widespread hydrogen economy [2]. Traditional storage techniques, like high-pressure gas or cryogenic liquid, exhibit inherent safety concerns or energy penalties, directing significant attention towards solid-state storage of hydrogen in hydrides [3].

Among a variety of hydrogen storage materials, magnesium (Mg) still stands out because of its exceptional gravimetric hydrogen storage capacity (7.6 wt% in $MgH_2$), high availability, and low price, making it a compelling candidate for practical applications [4]. Despite these advantages, the widespread implementation of magnesium-based hydrides is severely limited by two major obstacles: (i) the high thermodynamic stability of $MgH_2$, which necessitates elevated temperatures (above 573 K) for hydrogen release, and (ii) the sluggish kinetics of hydrogen absorption and desorption (sorption), which limit its operational practicality [5]. To overcome the thermodynamic barrier, other elements should be added to magnesium to weaken the hydrogen binding energy [6]. Although this strategy leads to the reduction of dehydrogenation temperature, it also lowers the hydrogen storage capacity [6]. Therefore, magnesium is basically considered a storage medium for applications where heating is possible [4,5].

Regarding the kinetic issue, the sluggish kinetics of hydrogen storage in magnesium are primarily ascribed to the poor hydrogen dissociation rate on the magnesium surface, slow metal/hydride interphase growth or the sluggish transport of hydrogen via the growing layer of magnesium hydride [4,5]. Over the recent decades, large attempts have been dedicated to overcoming these kinetic limitations [7,8]. Using catalysts is one common approach to address the kinetic problems of magnesium [7,8]. Another common and effective strategy involves the introduction of lattice imperfections and grain boundaries, which serve as rapid diffusion routes for hydrogen and provide preferential sites for the nucleation of hydride [9,10]. Severe plastic deformation techniques [11,12], like equal-channel angular pressing [13,14], high-pressure torsion (HPT) [15,16], intensive rolling [17,18] and fast forging [19,20], have proven particularly powerful in this regard, refining the microstructure of magnesium to the ultrafine-grained or nanocrystalline regime and thereby generating high densities of grain boundaries. This microstructural refinement alone has been reported to markedly improve hydrogen storage kinetics in pure magnesium [21,22], although low thermal stability is a common bottleneck of severely deformed ultrafine-grained materials [11,12].

A more promising approach to further activate hydrogen storage materials involves the creation of high densities of interphase boundaries [23,24]. Unlike grain boundaries, which separate grains of the same phase, interphase boundaries are interfaces between two distinct materials and are usually stable at high temperatures [25,26]. By introducing a second phase, interfaces can provide potent paths for hydrogen transport and also function as locations for heterogeneous nucleation of hydride [27,28]. Graphene, with a large surface area [29], superior thermal stability/conductivity [30], and known catalytic effects on hydrogen dissociation [31], is an ideal candidate as a secondary phase [32,33]. There have been some attempts to introduce light-weight graphene in magnesium using ball milling [34,35], but the magnesium powders processed by this method usually suffer from poor air resistance and must be handled in a glove box under a controlled atmosphere [36]. In contrast, bulk materials produced by severe plastic deformation are shown to be air resistant and keep their activity high even after long-term air exposure [37,38]. So far, there have been no attempts to introduce graphene in magnesium using severe plastic deformation to improve the kinetic features.

In this study, two strategies are synergistically combined in magnesium to dramatically improve the hydrogen storage kinetics: (i) creation of abundant graphene-magnesium interphase boundaries, and (ii) synthesis of bulk samples by severe plastic deformation. HPT is employed not merely as a grain refinement tool, but as a unique synthesis method to forcibly intermix magnesium and graphene. This approach generates a bulk nanocomposite structure with a high density of magnesium-graphene interphase boundaries. The results reveal that while HPT processing of pure magnesium improves kinetics through grain boundaries, the addition of graphene combined with HPT leads to a further, pronounced enhancement through interphase boundaries. Significantly, the activation energy for hydrogen desorption remains largely unchanged across all processed states; however, a dramatic improvement in kinetics is achieved through a substantial increase in the pre-exponential frequency factor, which correlates directly with the number of available nucleation locations and diffusion routes provided by the network of magnesium-graphene interphase boundaries. This work highlights the critical significance of interphase boundary engineering in designing advanced hydrogen storage materials.

## 2. Experimental Procedures

### 2.1. Materials and Processing

The starting materials were magnesium powder (>99% purity) and graphene. Three composite mixtures were prepared with graphene additions of 0, 2, 5 and 10 wt%. For each mixture, the powder blend was first dispersed in acetone and mixed thoroughly for 30 min to acquire a preliminary homogeneous distribution of graphene with magnesium. The resulting slurry was then dried at 333 K in an oven for 24 h to completely evaporate the acetone. Following drying, the mixed powders with a mass of about 100 mg were compacted into discs with a radius of 5 mm utilizing a hand press with a uniaxial pressure of 0.3 GPa at room temperature. These green compacts were subsequently processed by HPT, a method described in detail elsewhere [39,40]. HPT was carried out at ambient temperature under a quasi-hydrostatic pressure of 6 GPa. The sample was rotated at a rate of 1 rpm for a total of 20 turns, imposing a severe shear strain on the materials. For reference, discs of pure magnesium (99.9% purity) were prepared from an ingot and annealed for 1 h at 773 K.

### 2.2. Characterization

The microstructures of the samples were characterized using multiple techniques. Phase characterization was carried out utilizing X-ray diffraction (XRD) with Cu Kα radiation. The microstructure of the annealed sample was evaluated by optical microscopy (OM) after etching using a mixture of 5 vol% $HNO_3$ and 95 vol% $C_2H_5OH$. The overall morphology and graphene distribution were characterized utilizing scanning electron microscopy (SEM) with a voltage of 15 keV. Electron back-scatter diffraction (EBSD) was conducted on an HPT-processed sample without graphene after fine mechanical polishing, but well-polished surfaces could not be achieved from the samples containing graphene. The step size for EBSD was 46.5 nm and the electron energy was 20 keV. For detailed nanostructural observation, transmission electron microscopy (TEM) was utilized. TEM foils were primarily fabricated by electropolishing, following a procedure similar to that described earlier. Discs with a 1.5 mm radius were cut at 2-5 mm away from the sample center. The thickness of these discs was reduced to 0.15 mm by sandpaper grinding and further to a thin thickness for electron transparency using an electro-chemical polishing system under a voltage of 30 V at 263 K in a chemical mixture of 2 vol% $HClO_4$, 28 vol% $C_3H_5(OH)_3$ and 70 vol% $CH_3OH$ [22]. After electropolishing, the TEM samples were cleaned using ion milling conducted at an angle of ±5º with an energy of 5 keV for 15 min.

### 2.3. Hydrogen Storage Property Measurements

The hydrogen storage features were evaluated utilizing a Sieverts-type machine. Before the hydrogenation test, samples were crushed for 30 min using a pestle and mortar. Crushing was conducted in an air atmosphere to clarify the air resistance of these materials. After crushing, approximately 60 mg of the samples were used for absorption and desorption kinetic measurements. Before hydrogenation testing, no special activation treatment was conducted, but samples were evacuated for 2 h to eliminate possible moisture. Hydrogen absorption (hydrogenation) kinetics were measured under a primary hydrogen pressure of 4.03 MPa at 623 K. Desorption (dehydrogenation) kinetics were evaluated under a primary back pressure of 0.01 MPa at 623 K. To understand the hydrogenation products, XRD was performed after hydrogenation. Moreover, to determine the kinetic parameters, thermal analysis was conducted on hydrogenated samples. Concurrent thermogravimetry (TG) and differential scanning calorimetry (DSC) were conducted on approximately 12 mg of sample under a high-purity argon gas in the temperature range of 303-873 K. For calculating the activation energy of hydrogen storage via the Kissinger approach [41], TG-DSC measurements were conducted at four heating speeds of 2.5, 5, 10 and 20 K/min.

## 3. Results

### 3.1. Structural and Microstructural Evolution

Fig. 1 presents the XRD patterns of the annealed pure magnesium and the HPT-processed magnesium powder mixed with 0, 2, 5 and 10 wt% of graphene. The patterns for all samples primarily show peaks corresponding to the hexagonal close-packed (HCP) structure of magnesium. For samples containing graphene, a weak peak appears at 26.5º, which corresponds to the (002) atomic plane of graphene. No peaks corresponding to magnesium carbide or other magnesium-graphene reaction products are detected, suggesting that graphene remains as a distinct phase within the magnesium matrix. Another feature is that for the annealed sample, the pyramidal (101) plane has the highest peak intensity, but the peak intensity for the basal (002) plane increases by HPT treatment, and it becomes more prominent by the addition of graphene.

The impact of HPT on grains was further examined using OM for the annealed sample and using EBSD for HPT-processed magnesium without graphene addition, as illustrated in Fig. 2. For the annealed sample (Fig. 2a), grain sizes are in the order of 1 mm. For the HPT-treated pure magnesium (Fig. 2b), the microstructure contains refined grains with a mean size of 850 nm. This grain size reasonably agrees with earlier publications on HPT processing of pure magnesium [22,42]. Moreover, EBSD analysis suggests that 70% of grain boundaries have high misorientation angles over 15º, while low-angle grain boundaries with misorientations of 2-15º account for 30% of boundaries. Grain refinement by HPT is a general observation not only after HPT [39,40], but also after any other severe plastic deformation methods [13-20].

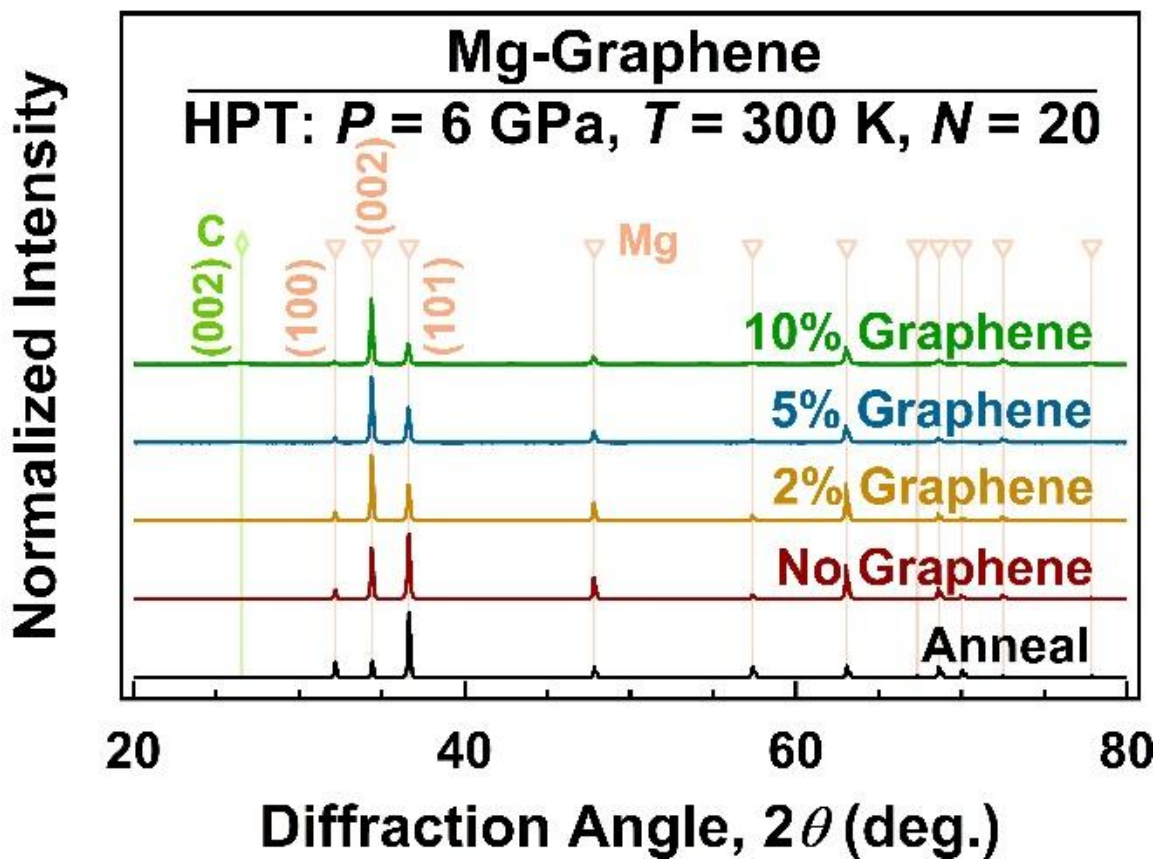

Fig. 1. Mixing of magnesium with graphene by HPT with no chemical reactions. XRD profiles of magnesium mixed with 0, 2, 5 and 10 wt% of graphene by HPT as well as annealed pure magnesium.

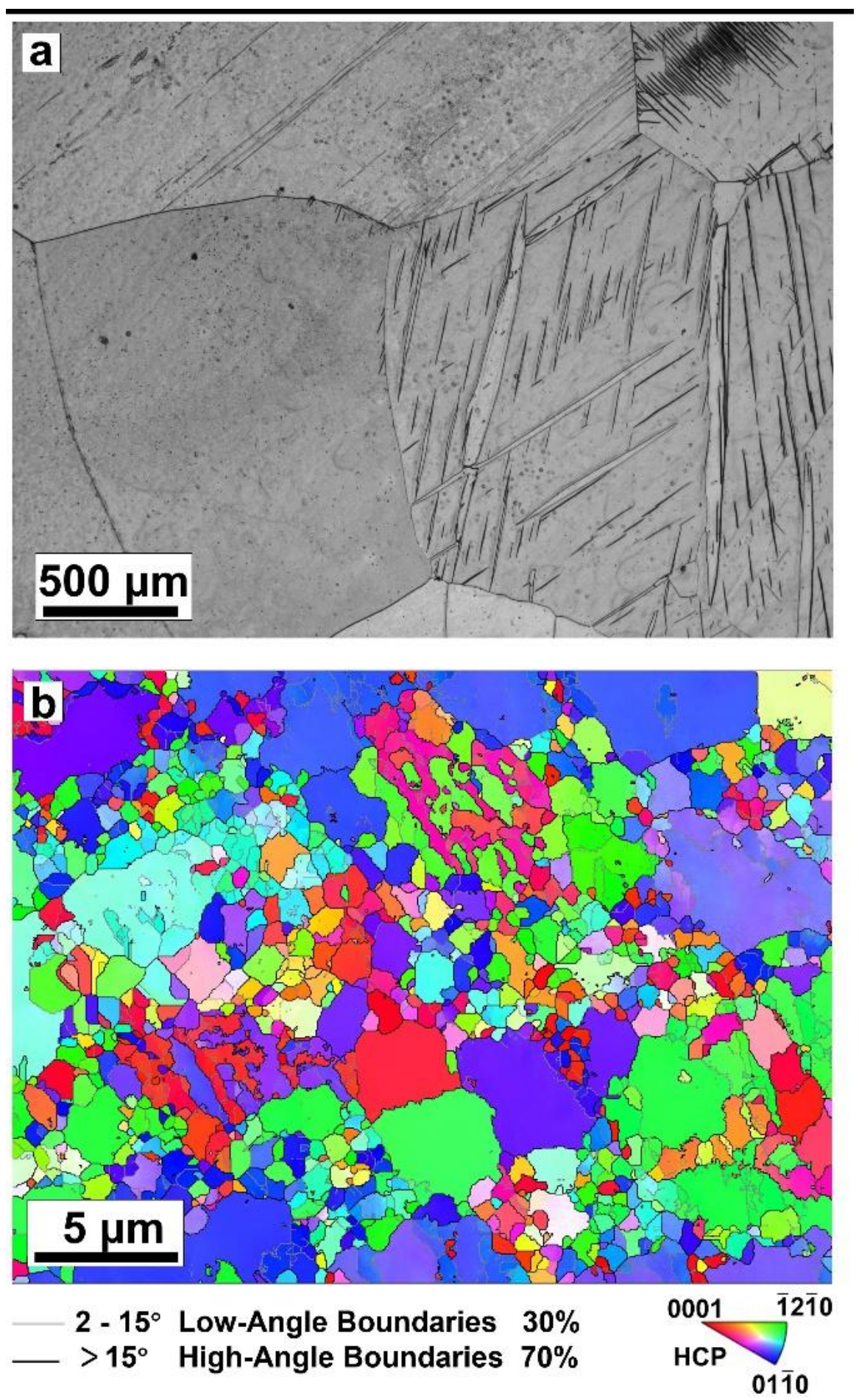

Fig. 2. Formation of large fractions of high-angle grain boundaries by HPT. (a) OM image of annealed magnesium and (b) EBSD orientation map of HPT-processed magnesium.

To examine the impact of graphene on microstructural evolution, the graphene-containing composites were first studied by SEM and energy dispersive X-ray spectroscopy (EDS). Such examinations were carried out on both the disc surface and cross-section, as displayed in Fig. 3. Examination of the graphene dark phase in magnesium bright matrix indicates that the density of carbon-rich regions increases proportionally with the nominal graphene content. The carbon EDS maps for the 2, 5 and 10% graphene samples demonstrate a bimodal dispersion of graphene throughout the magnesium matrix, including fine particles and the agglomerated ones. The agglomeration of graphene is more significant in the sample containing 10% graphene.

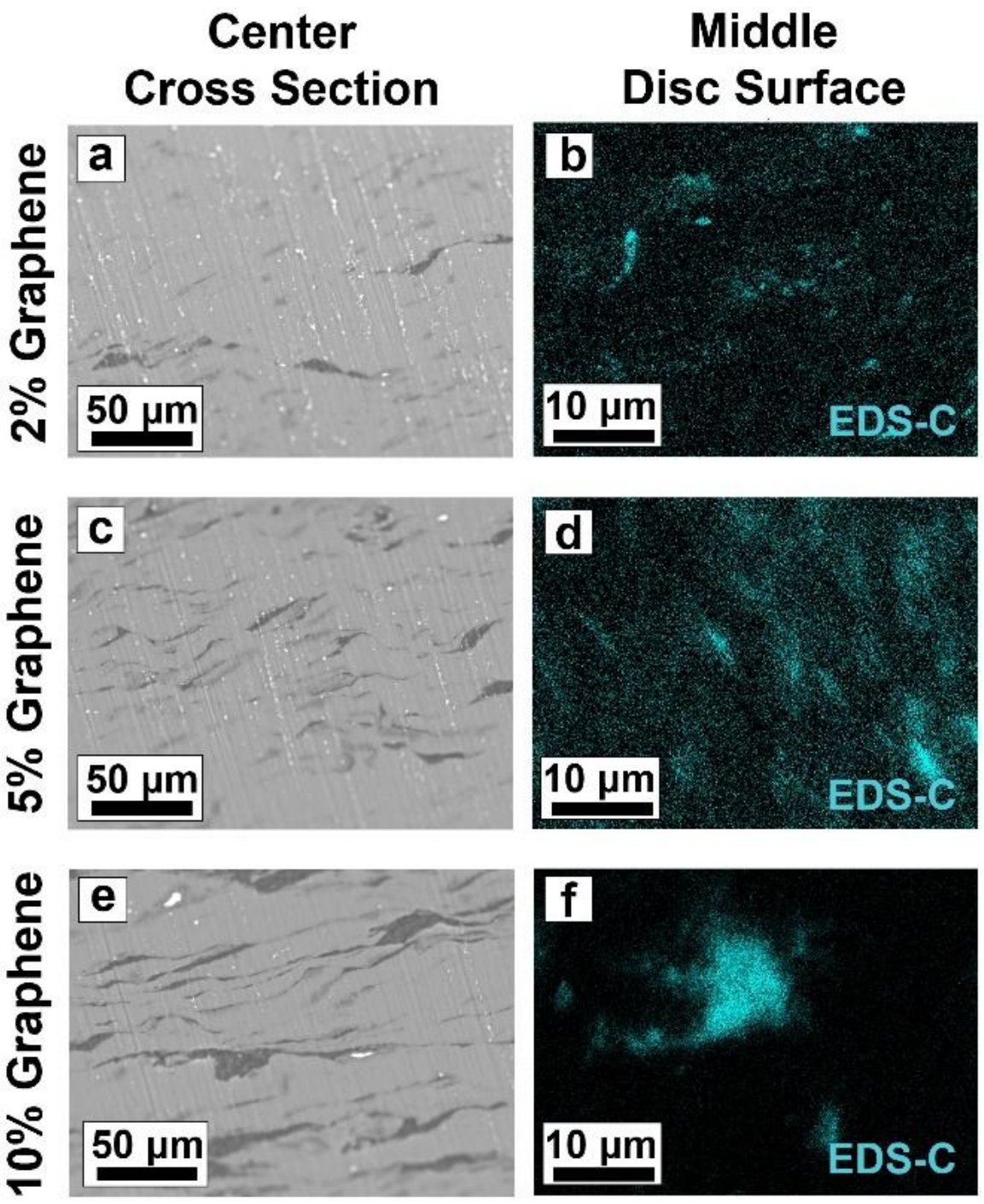

Fig. 3. Distribution of graphene in magnesium by HPT. (a, c, e) SEM images taken from disc center at cross-sectional view and (b, d, f) EDS maps of carbon taken from middle of discs at top view for magnesium mixed with (a, b) 2, (c, d) 5 and (e, f) 10 wt% of graphene by HPT

The TEM observations provide a more detailed view of the microstructure and the influence of graphene on the microstructure. Fig. 4a-c shows bright-field images, dark-field images and relevant selected area electron diffraction (SAED) of annealed pure magnesium, while Fig. 4d-f and 4g-l show similar results for the HPT-processed pure magnesium and the sample containing 5% graphene, respectively. Consistent with the OM and EBSD observations, the annealed magnesium has coarse grains, while the HPT-processed pure magnesium exhibits a refined grain structure with sizes at the submicrometer level. The distorted structure of grains in both dark- and bright-field images after HPT processing suggests the existence of dislocations and low-angle boundaries in the microstructure. The sample containing 5% graphene displays a bimodal microstructure. In some regions, as shown in Fig. 4g-i, its structure contains submicrometer grains with sizes up to 500 nm. In some regions, as shown in Fig. 4j-l, it contains nanograins with grain sizes down to 10 nm. The significantly finer microstructure in this region can also be observed from a complete ring form of the SAED profile. More pronounced grain refinement in the presence of graphene should be because of the effect of this second phase on pinning grain boundaries [11,12].

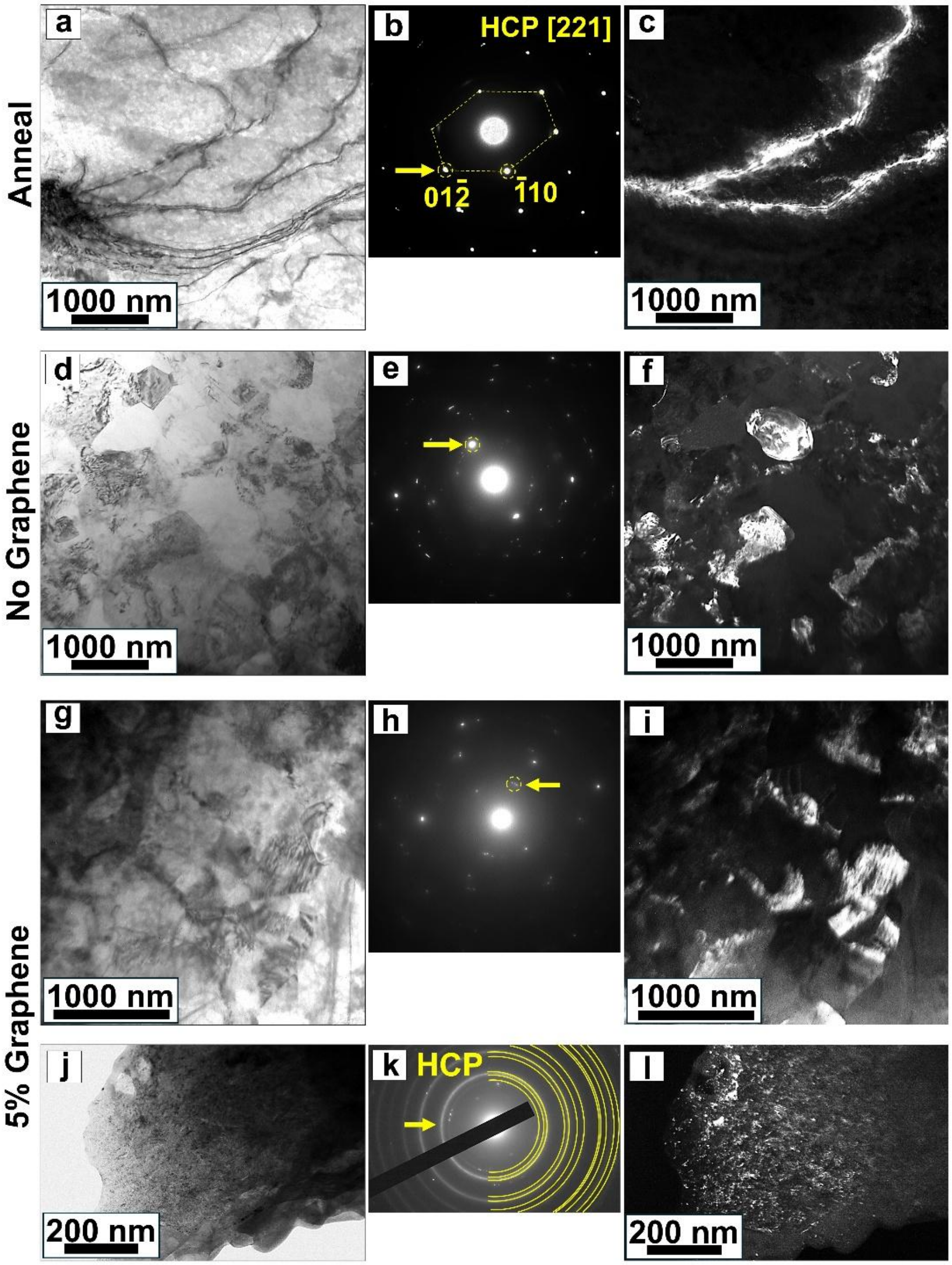

Fig. 4. Grain refinement by application of HPT to magnesium-graphene composites. TEM (a, d, g, j) bright-field micrographs, (b, e, h, k) SAED analyses and (c, f, i, l) dark-field micrographs for (a-c) annealed pure magnesium, (d-f) HPT-processed pure magnesium and (g-l) magnesium mixed with 5 wt% of graphene by HPT. Dark-field micrographs were generated utilizing diffracted beams indicated by arrows in SAED.

High-resolution TEM images and relevant fast Fourier transform (FFT) patterns in Fig. 5 provide further insight into the nanostructure and interface structure of the samples. Fig. 5a-c shows the lattice fringes of the HCP magnesium in the HPT-treated pure sample, indicating the existence of dislocations within grains. The formation of dislocations is a general feature of HPT-treated metals, but their density is dependent on the purity and melting point of the metals [43]. High-resolution TEM images for the magnesium with 5% graphene composite treated by HPT are presented in Fig. 5d and 5e. The interface between the graphene flake and the magnesium matrix can be observed in the composite.

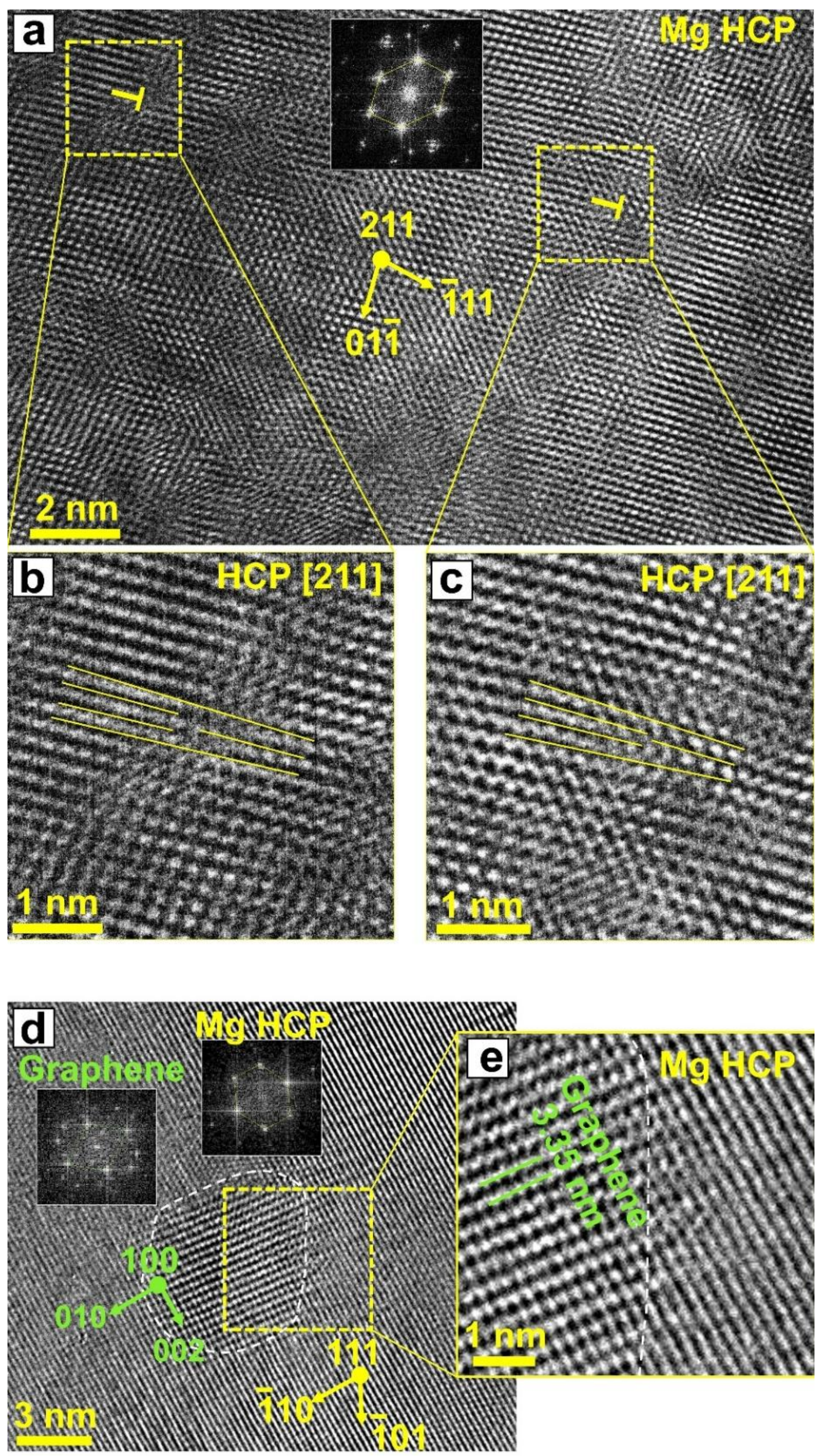

Fig. 5. Formation of dislocation and magnesium-graphene interphase boundaries by HPT. TEM high-resolution images and relevant FFT analyses for (a-c) HPT-processed pure magnesium and (d, e) magnesium mixed with 5 wt% of graphene by HPT. (b) and (c) are magnified images of dislocations shown in (a) and (e) is a magnified view of interphase shown in (d).

### 3.2. Hydrogen Storage Properties

The hydrogen absorption kinetics of the material were evaluated at 623 K with a primary hydrogen pressure of 4.03 MPa, as depicted in Fig. 6a. The annealed pure magnesium shows extremely sluggish absorption kinetics. HPT processing of pure magnesium significantly accelerates the absorption rate, reaching a higher hydrogen capacity in a shorter time. However, the magnesium-graphene composites exhibit a further, remarkable enhancement in absorption kinetics. The absorption amount decreases slightly with increasing the graphene content to 10%. The stored hydrogen in magnesium graphene composites is close to the theoretical gravimetric capacity of magnesium, 7.6% [4-6].

To confirm the extent of hydrogenation for different samples, XRD analysis was carried out on the materials following the hydrogen absorption tests, as illustrated in Fig. 6b. The patterns for all hydrogenated samples clearly show the generation of β-$MgH_2$. For the annealed sample, an intense peak for the remaining magnesium phase is still visible. The intensity of these peaks decreases by HPT processing of pure magnesium, and they totally disappear in the composites containing graphene. These results agree with the hydrogen capacities achieved in the kinetic measurements in Fig. 6a, confirming that the magnesium-graphene composites are fully hydrogenated due to fast kinetics [7-10].

A trend similar to the absorption kinetics is observed for the desorption kinetics, as presented in Fig. 6c. The dehydrogenation rate of the HPT-processed pure magnesium is faster than that of the annealed sample. The magnesium-graphene composites again show superior performance, desorbing hydrogen quickly. By considering both absorption and desorption results, the sample with 5% graphene can be considered the best hydrogen storage material among the five investigated materials.

The thermal desorption features of the hydrogenated materials were further investigated using DSC and TG. Fig. 7a and 7b show representative DSC and TG curves, respectively, for the hydrogenated samples at a heating speed of 5 K/min. The DSC profiles for all samples exhibit a strong endothermic peak arising from the decomposition of $MgH_2$. Compared to annealed magnesium, the onset temperature for the dehydrogenation peak for HPT-processed pure magnesium is shifted to a lower temperature, in agreement with an earlier study [44]. The dehydrogenation temperature for the magnesium-graphene composites is shifted to a notably lower temperature, and the highest shift is detected for the sample containing 5% of graphene. This shift indicates that the presence of an optimum amount of graphene makes the dehydrogenation kinetically easier, but it does not indicate any changes in the thermodynamics [2,3]. The TG curves corroborate this, showing that the onset temperature for weight loss due to hydrogen release decreases after HPT treatment, and it is the minimum for the composite containing 5% graphene. It should be noted that the onset temperatures in Fig. 7 were estimated from the intersection of baselines and tangent lines of the steepest slopes after transition.

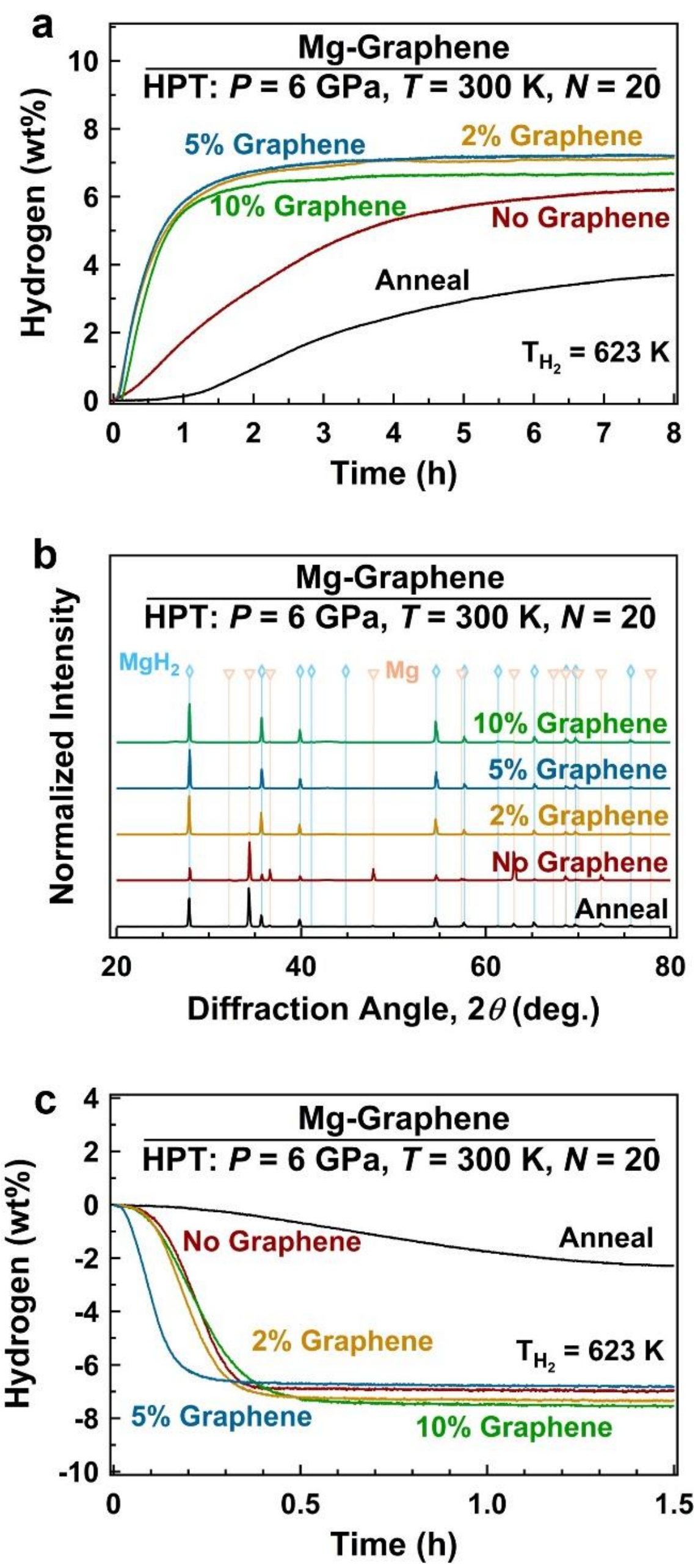

Fig. 6. Fast hydrogenation and dehydrogenation kinetics in magnesium mixed with graphene by HPT. (a) Hydrogen absorption versus time, (b) XRD profiles after hydrogenation and (c) hydrogen desorption versus time for magnesium mixed with 0, 2, 5 and 10 wt% of graphene by HPT as well as annealed pure magnesium. Weight percentage of absorbed hydrogen is based on the mass of only magnesium and without considering the mass of graphene.

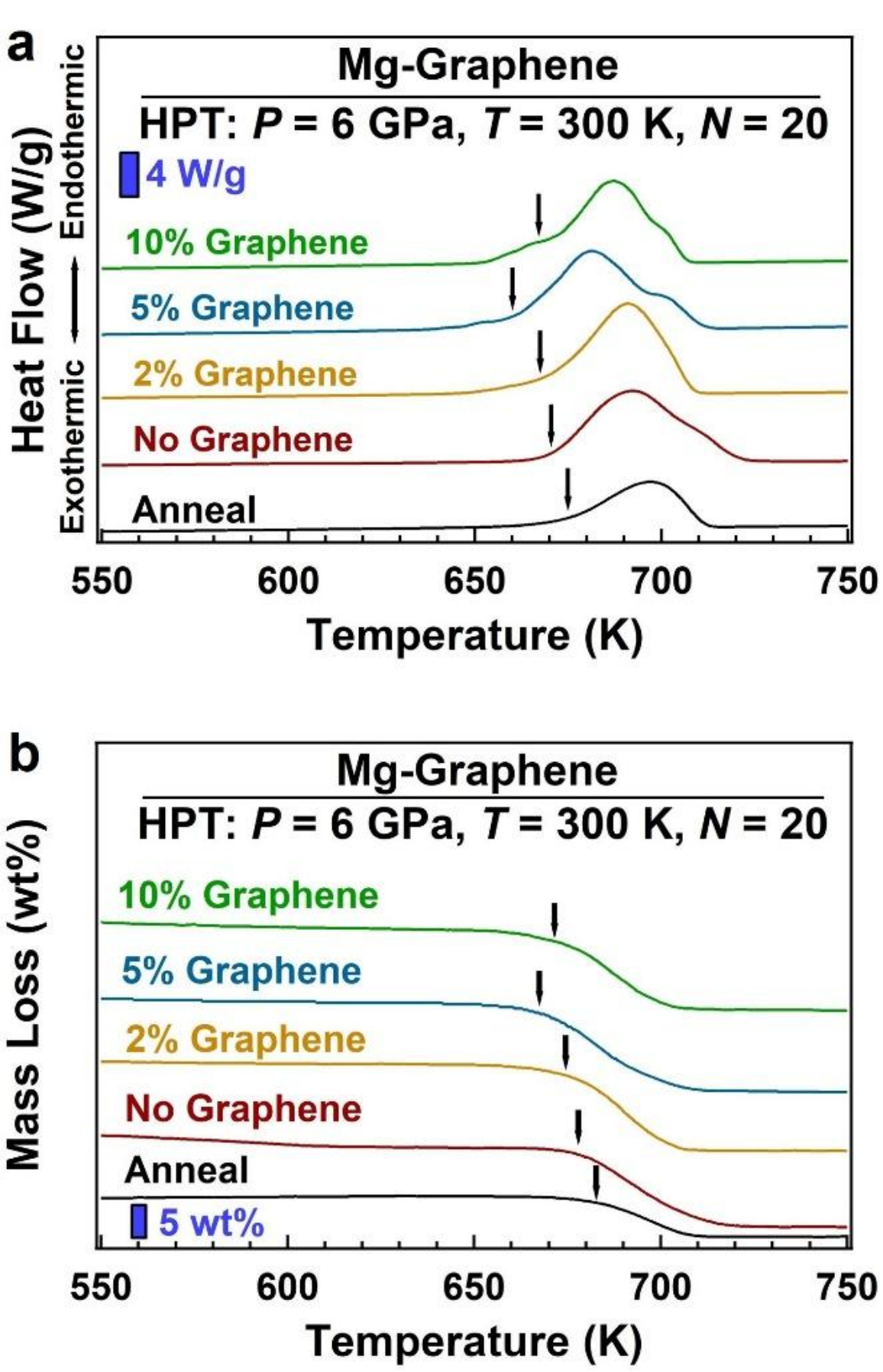

Fig. 7. Reduction of onset temperature of dehydrogenation in magnesium mixed with graphene by HPT. (a) Heat flow measured by DSC and (b) mass loss measured by TG versus time for magnesium mixed with 0, 2, 5 and 10 wt% of graphene by HPT as well as annealed pure magnesium. Arrows show onset temperatures.

**4. Discussion**

The results of the present investigation exhibit that the application of HPT to add graphene to magnesium leads to air resistance and a dramatic enhancement in the hydrogen storage kinetics of magnesium. While HPT alone is a well-established method for producing bulk nanostructured materials with high defect densities [39,40], its application here serves a dual purpose: first, it refines the magnesium matrix to generate a high density of grain boundaries [21,22]; and second, it acts as a synthesis tool [37,38] to forcibly intermix immiscible magnesium and graphene. Thereby, the process enables synthesizing a bulk nanocomposite with an abundant network of magnesium-graphene interphase boundaries. The following discussion addresses three key aspects of the current investigation: (i) the unique role of interphase boundaries as thermally stable defects for hydrogen transport and hydride nucleation even after air exposure, (ii) the changes in the hydrogenation rate-controlling mechanism by introducing grain boundaries and interphases, and (iii) the changes of the activation energy and frequency factor by introducing grain boundaries and interphase boundaries.

**4.1. Interphase Boundaries as Thermally Stable Hydrogen Pathways**

The microstructural characterization (Fig. 2-5) clearly shows that HPT processing of magnesium with graphene results in a significantly finer grain size than HPT processing of pure magnesium alone. This indicates that graphene acts as a powerful grain refiner during severe plastic deformation, pinning dynamic recrystallization and preventing grain boundary migration [45]. More importantly, the EDS and TEM observations confirm the existence of magnesium-graphene interphase boundaries. These interfaces are distinct from conventional grain boundaries in two critical ways. First, they are interphase boundaries between two chemically dissimilar materials, which inherently possess a different atomic structure and bonding environment compared to magnesium-magnesium grain boundaries. Second, and most significantly for hydrogen storage applications, these interphase boundaries exhibit high thermal stability [25,26] because graphene pins the magnesium crystals. While grain boundaries in nanocrystalline magnesium can migrate and coarsen at elevated temperatures during hydrogen cycling, the magnesium-graphene interfaces remain stable due to the immiscibility and high thermal stability of graphene with a large surface area [29-31]. This thermal stability ensures that the fast diffusion pathways and heterogeneous nucleation sites provided by these interfaces are preserved [32-35]. The schematic in Fig. 8 illustrates this concept, showing hydrogen dissociating on the surface and rapidly diffusing along the thermally stable magnesium-graphene interphase boundaries to reach the magnesium matrix for heterogeneous nucleation of $MgH_2$ even after air exposure.

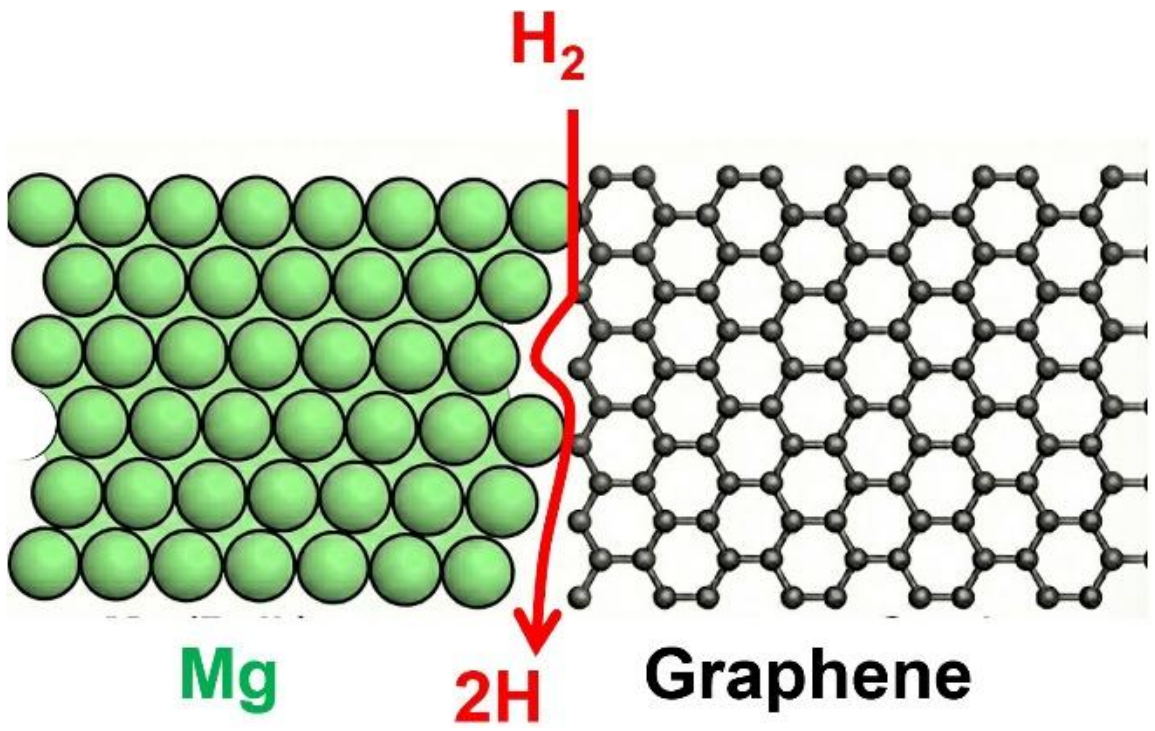

Fig. 8. Schematic representation of magnesium-graphene interphase boundaries showing their importance for hydrogen diffusion from surface to bulk.

### 4.2. Effect of Boundaries on Hydrogenation Mechanisms

The kinetic modeling of the absorption data can provide profound insight into the rate-controlling mechanism for hydrogen storage [46]. Several models have been suggested in the literature regarding the hydrogenation kinetics and rate-controlling mechanisms, including one-dimensional diffusion [47], two-dimensional diffusion [48], three-dimensional diffusion [49-52], random nucleation and growth [53,54], two-dimensional interfacial reaction [47], and three-dimensional interfacial reaction [48]. There are also models based on zero-order kinetics and first-order kinetics [47,48]. A summary of the most popular models is given in Table 1, where $\xi$ denotes the fraction of absorbed hydrogen, $t$ denotes hydrogenation time and $k$ denotes the reaction kinetic constant. The evaluation of the best-fit models for the current study reveals a systematic transition in the rate-limiting stage, from interface-controlled growth to diffusion-controlled growth, driven by the increasing density of sites for heterogeneous nucleation on grain boundaries and interphase boundaries, as shown in Fig. 9.

Table 1. Kinetic models for hydrogen storage by different rate-controlling mechanisms.

| Model | Mechanism | Ref. |
|---|---|---|
| $\xi^2 = kt$ | One-dimensional diffusion | [47] |
| $(1-\xi)\ln(1-\xi) + \xi = kt$ | Two-dimensional diffusion | [48] |
| $[1 - (1-\xi)^{1/3}]^2 = kt$ | Three-dimensional diffusion (Jander model) | [49] |
| $[(1/(1-\xi))^{1/3} - 1]^2 = kt$ | Three-dimensional diffusion (modified Jander model) | [50] |
| $1 - 2\xi/3 - (1-\xi)^{2/3} = kt$ | Three-dimensional diffusion (Gisling–Braunshtein model) | [51] |
| $[1 - (1-\xi)^{1/3}]^2 = kt$ | Three-dimensional diffusion (Kroger-Ziegler model) | [52] |
| $[-\ln(1-\xi)]^{1/2} = kt$ | Random nucleation and growth (Avrami-Erofeev model) | [53] |
| $[-\ln(1-\xi)]^{1/3} = kt$ | Random nucleation and growth (Avrami-Erofeev model) | [54] |
| $1 - (1-\xi)^{1/2} = kt$ | Two-dimensional interfacial reaction | [47] |
| $1 - (1-\xi)^{1/3} = kt$ | Three-dimensional interfacial reaction | [48] |
| $\xi = kt$ | Zero-order kinetics | [47] |
| $-\ln(1-\xi) = kt$ | First-order kinetics | [48] |

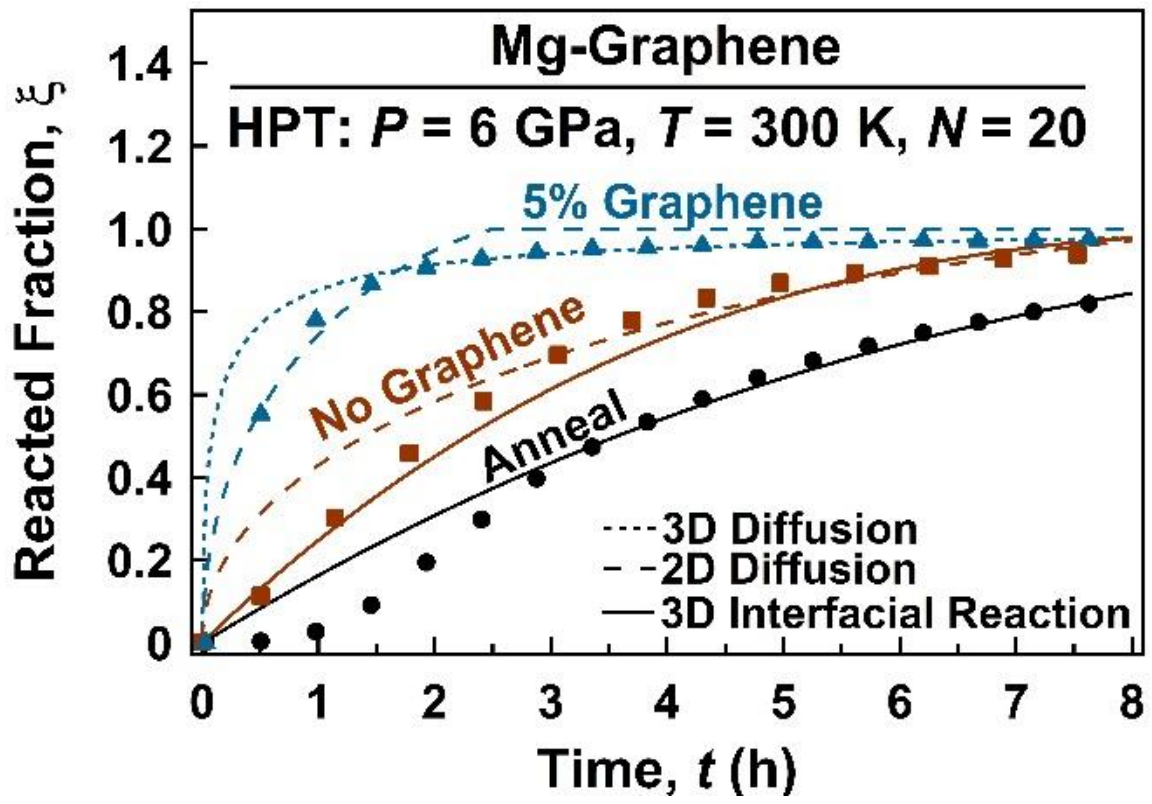

Fig. 9. Transition of rate-controlling mechanism from three-dimensional interfacial reaction to two-dimensional diffusion by introduction of magnesium-graphene interphase boundaries. Reacted fractions obtained experimentally by hydrogenation kinetic measurements and calculated by three different kinetic models for annealed pure magnesium, HPT-processed pure magnesium and magnesium mixed with 5 wt% of graphene by HPT.

For the annealed pure magnesium, the best fit was obtained with a three-dimensional interfacial reaction model ($R^2$ = 0.992). This model supposes that the kinetics is dominated by the movement of the Mg/$MgH_2$ phase boundary interfaces [48]. In coarse-grained magnesium with few defects, the hydride forms a continuous shell around the magnesium particles, and the reaction proceeds inward, with the rate limited by the speed at which this interface can advance. This is the classical behavior of inactivated magnesium, where hydride growth is geometrically constrained by the inward progression of a single reaction front [55].

After HPT processing of pure magnesium, the kinetic analysis revealed a more complex situation. The absorption data for this sample were equally well described by two different models: (i) the three-dimensional interfacial reaction model ($R^2$ = 0.990) and (ii) a two-dimensional diffusion model ($R^2$ = 0.991). The coexistence of two statistically equivalent fits suggests that the HPT-processed pure magnesium exists in a microstructural transition state. While the material still contains some regions where the interfacial movement is rate-limiting, the introduction of a high density of grain boundaries has also created fast diffusion pathways. The two-dimensional diffusion fit implies that hydrogen transport along grain boundaries and the subsequent nucleation of hydride at these boundaries have become a significant competing mechanism [48].

Remarkably, for the sample of magnesium with 5% graphene, the two-dimensional diffusion model ($R^2 = 0.990$) and the three-dimensional diffusion model, so-called modified Jander model ($R^2 = 0.989$), provided the best fit. The interfacial reaction model no longer provided a satisfactory description of the kinetics in the presence of graphene. This result indicates a fundamental shift in the rate-controlling mechanism. At first glance, a diffusion-controlled model might be misinterpreted as implying slower kinetics. However, as shown in Fig. 6a, the composite absorbs hydrogen the fastest. The explanation lies in the unique microstructural architecture created by HPT processing with graphene. In the magnesium-graphene composite, the hydrogen dissociation and absorption occur readily at the abundant interphase boundaries [32-35]. These interfaces should also function as powerful heterogeneous nucleation places for the hydride phase in hydrogenation and for the metal phase in dehydrogenation [25,27,28]. Because nucleation occurs simultaneously at thousands of interphase sites throughout the volume [56,57], the growing $MgH_2$ regions quickly impinge upon one another. Once this surrounding magnesium is transformed into a continuous or semi-continuous $MgH_2$ layer, further hydrogenation of the remaining magnesium interior requires hydrogen to diffuse through this growing hydride product layer. The better fit to a two-dimensional model, rather than a three-dimensional one, suggests that the diffusion geometry is not spherically symmetric from a point source; instead, it implies that hydrogen diffusion occurs preferentially along the interphase boundaries.

### 4.3. Effect of Boundaries on Activation Energy and Frequency Factor

The Kissinger analysis, applied to DSC measurements at different heating rates, can provide extra information regarding the kinetics [41].

$$\phi/T^2 = A\exp(-Q/RT) \quad (1)$$

where $\phi$ denotes the speed of heating, $T$ denotes the absolute temperature at the peak of DSC, $A$ denotes the frequency factor whose increase enhances the kinetics, $Q$ denotes the activation energy whose increase diminishes the kinetics, and $R$ denotes the gas constant. The Kissinger plots, i.e. $\ln(\phi/T^2)$ plotted against $1/T$, are presented in Fig. 10.

The results show that the calculated activation energy for hydrogen desorption remains essentially unchanged at about 145±2 kJ/mol across annealed and HPT-processed samples, with or without graphene. This constant activation energy, which is slightly higher than earlier reports [58,59], indicates that the energy barrier for breaking Mg-H and releasing the hydrogen from the material does not significantly alter by the presence of grain boundaries or interphase boundaries. While the activation energy remains unchanged, the dramatic improvement in kinetics in the presence of interphase boundaries is due to an increase in the pre-exponential frequency factor (intercept of plots), which increases by HPT and further increases in the presence of graphene. The frequency factor indicates the frequency of attempts to overcome the energy barrier and is proportional to the number of active sites available for the reaction [41,60]. This increase in the frequency factor can be directly correlated with the microstructural features observed in Fig. 2-5. HPT processing of pure magnesium increases the frequency factor by creating a high fraction of grain boundaries, which function as locations for dehydrogenation. The significance of these HPT-induced grain boundaries on hydrogen storage was proved in earlier studies [22,61]. The addition of graphene and the subsequent creation of a high density of interphase boundaries dramatically increase the number of successful attempts to overcome the energy barrier. As a result, a much faster overall reaction rate is achieved, even though the fundamental energy barrier for each event remains the same [60]. Although the current results provide a solution to address the kinetic issues, modification of the thermodynamic parameters remains an open issue that should be addressed by chemical modification through a combination of theoretical calculations, machine learning and experiments [62,63].

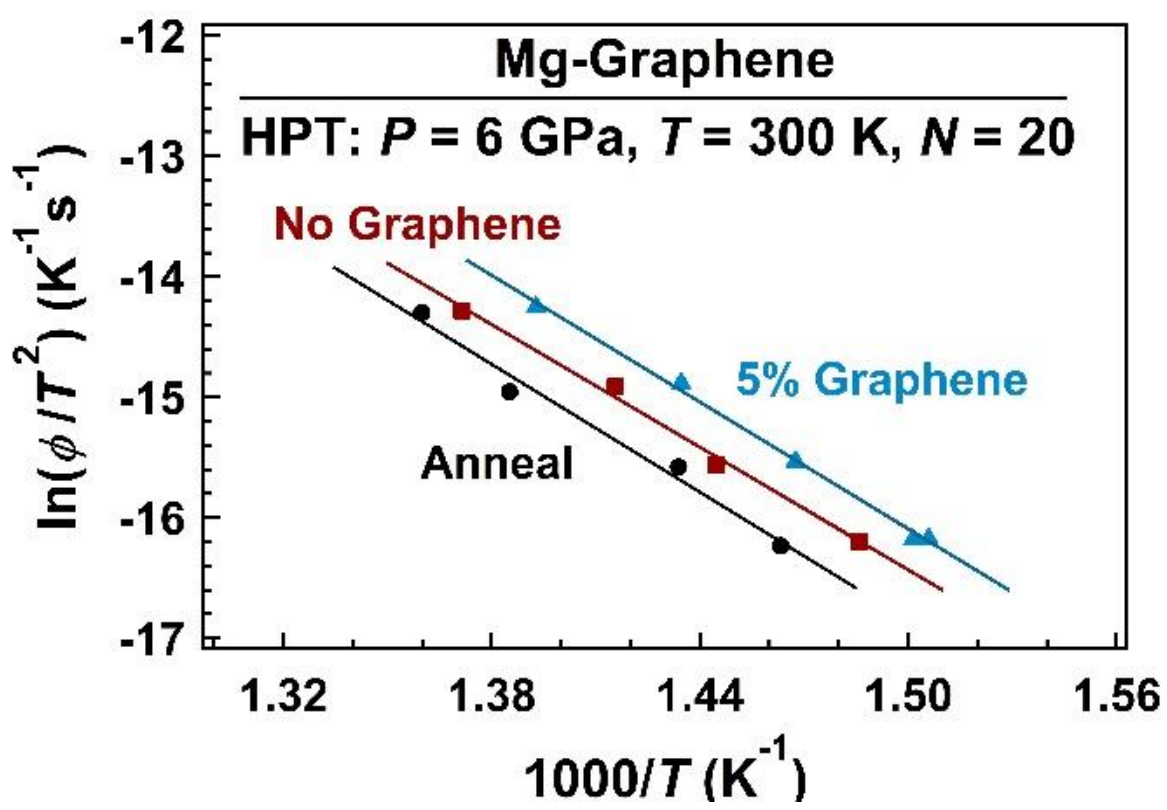

Fig. 10. No change in activation energy and enhancement of pre-exponential frequency factor by introduction of magnesium-graphene interphase boundaries. Kissinger plots ($\phi$: Heating rate, $T$: temperature) for annealed pure magnesium, HPT-processed pure magnesium and magnesium mixed with 5 wt% of graphene by HPT.

## 5. Conclusions

This study systematically investigates the effect of magnesium-graphene interphase boundaries, created by high-pressure torsion (HPT), on the hydrogen storage kinetics of magnesium. The following conclusions are drawn.

1. HPT treatment of pure magnesium refines grain sizes from ~1 mm to 850 nm, with 70% of grain boundaries exhibiting high misorientation angles. The addition of graphene during HPT results in a bimodal nanostructure with grain sizes ranging from 10 to 500 nm, where graphene flakes are dispersed in magnesium and form interphase boundaries.
2. Magnesium-graphene composites processed by HPT exhibit superior hydrogen absorption and desorption kinetics at 623 K compared to both annealed and HPT-processed pure magnesium, while maintaining high air resistance. A composite with 5 wt% graphene shows the best overall performance, achieving full hydrogenation close to the theoretical capacity of magnesium.
3. The rate-controlling mechanism for hydrogen absorption undergoes a systematic transition with grain and interphase boundary introduction. Coarse-grained annealed magnesium is controlled by a three-dimensional interfacial reaction. HPT-processed pure magnesium exhibits a mixed regime, where both interfacial reaction and two-dimensional diffusion contribute. For the magnesium-graphene composite, the mechanism shifts to two-dimensional diffusion control.
4. Kissinger analysis reveals that the activation energy for hydrogen desorption remains essentially unchanged at 145±2 kJ/mol, regardless of the presence of grain boundaries or interphase boundaries. However, the pre-exponential frequency factor increases significantly with the generation of interphase boundaries. This increase correlates directly with the higher density of active sites for hydrogen diffusion and heterogeneous nucleation provided by interphases.

5. This work demonstrates that interphase boundary engineering by severe plastic deformation is a potent strategy for raising hydrogen storage kinetics in magnesium-based materials. The approach combines the benefits of grain refinement with the thermal stability of interphases with carbon-based materials, proposing a path for developing hydrogen storage materials with improved kinetics and air resistance.

**CRediT Authorship Contribution Statement**

All authors: Conceptualization, Methodology, Investigation, Validation, Writing – review & editing.

**Declaration of Competing Interest**

The authors declare no competing financial interests or personal relationships that could affect the results presented in this article.

**Acknowledgments**

This investigation is supported in part by the Japan Light Metal Educational Foundation, in part by the Magnesium Research Center of Kumamoto University, and in part by the Japan Science and Technology Agency through the ASPIRE Project (JPMJAP2332).

**Data Availability**

The data can be shared upon request from the corresponding author.